\documentclass[acmsmall]{acmart}

\usepackage{listings}
\usepackage{caption} % used in minipage
\usepackage{multirow}
\usepackage{xspace}
\usepackage{array}

\usepackage{amsmath}
\usepackage{algorithm}
\usepackage{algpseudocode}
\algnewcommand{\LeftComment}[1]{\State \(\triangleright\) #1}  % See https://tex.stackexchange.com/questions/184827/how-to-left-justify-comments-to-indent-same-as-other-text-in-algorithmic

\newcommand{\myrq}[1]{RQ{#1}}
\newcommand{\inlinecode}[1]{\texttt{\footnotesize\selectfont #1}}
\newcommand{\toolname}{\textsc{Restor}\xspace}
\newcommand{\fulltoolname}{\textbf{R}einforcement \textbf{E}nhanced \textbf{S}ingle-\textbf{T}raffic \textbf{O}racle generator for \textbf{R}EST APIs} % Reinforcement Enhanced Single-Traffic Oracle generator for REST APIs
\newcommand{\companyname}{ByteDance\xspace}

\newcommand{\baseModelFull}{Doubao-Seed-1.6-flash\xspace}
\newcommand{\baseModelShort}{Doubao-flash\xspace}

\usepackage{xcolor}
\usepackage[most]{tcolorbox}
\definecolor{LightGray}{rgb}{0.97,0.97,0.97}
\definecolor{DarkGray}{rgb}{0.25,0.25,0.25}

\usepackage[most]{tcolorbox}

\newtcolorbox{promptbox}[1]{
    enhanced,
    breakable,
    colback=white,
    colframe=black,        % Pure black border
    boxrule=1.5pt,         % Thicker, bold border matching your image
    arc=0pt,               % Sharp corners
    outer arc=0pt,
    colbacktitle=black,    % Solid black title bar
    coltitle=white,        % White title text
    fonttitle=\bfseries,   % Bold title text
    title={#1},            % The title argument
    pad at break=1mm,
    left=6pt, right=6pt, top=6pt, bottom=6pt % Slightly more inner padding
}

\tcbset{
    hlstyle/.style={
        on line, 
        boxsep=0pt, left=2pt, right=2pt, top=2pt, bottom=2pt, 
        arc=1pt, 
        colframe=#1, colback=#1
    }
}

\newcommand{\hlpink}[1]{\tcbox[hlstyle=red!10]{\textbf{#1}}}
\newcommand{\hlyellow}[1]{\tcbox[hlstyle=orange!15]{\textbf{#1}}}

\AtBeginDocument{%
  }

\setcopyright{cc}
\setcctype{by}
\acmDOI{10.1145/3832185}
\acmYear{2026}
\acmJournal{PACMSE}
\acmVolume{3}
\acmNumber{ISSTA}
\acmArticle{ISSTA094}
\acmMonth{10}
\acmSubmissionID{issta26main-p852-p}
\received{2026-01-30}
\received[accepted]{2026-04-16}

\begin{document}

%%
%% The "title" command has an optional parameter,
%% allowing the author to define a "short title" to be used in page headers.
\title{RESTOR: Automated Test Oracle Generation for RESTful APIs via Reinforcement Learning}

%%
%% The "author" command and its associated commands are used to define
%% the authors and their affiliations.
%% Of note is the shared affiliation of the first two authors, and the
%% "authornote" and "authornotemark" commands
%% used to denote shared contribution to the research.
\author{Xun Zhou}
\orcid{0009-0000-2929-416X}
\affiliation{%
  \institution{Fudan University}
  \city{Shanghai}
  \country{China}
}
\email{xunzhou24@m.fudan.edu.cn}

\author{Zhen Dong}
\authornote{Corresponding Author.}
\orcid{0009-0009-1193-0696}
\affiliation{%
  \institution{Fudan University}
  \city{Shanghai}
  \country{China}
}
\email{zhendong@fudan.edu.cn}

\author{Mingyu Ren}
\orcid{0009-0001-7656-9928}
\affiliation{%
  \institution{Fudan University}
  \city{Shanghai}
  \country{China}
}
\email{myren25@m.fudan.edu.cn}

\author{Qiang Li}
\orcid{0009-0003-4494-146X}
\affiliation{%
  \institution{ByteDance}
  \city{Shanghai}
  \country{China}
}
\email{liqiang.leo@bytedance.com}

\author{JunJie Li}
\orcid{0009-0003-4391-7492}
\affiliation{%
  \institution{ByteDance}
  \city{Shanghai}
  \country{China}
}
\email{wells.li@bytedance.com}

\author{Sifan Wang}
\orcid{0009-0005-4415-556X}
\affiliation{%
  \institution{ByteDance}
  \city{Shanghai}
  \country{China}
}
\email{wangsifan.28@bytedance.com}

\author{Xiaolong Yu}
\orcid{0000-0001-6505-8914}
\affiliation{%
  \institution{ByteDance}
  \city{Shanghai}
  \country{China}
}
\email{yuxiaolong.1@bytedance.com}

\author{Chaofeng Sha}
\orcid{0009-0004-4195-0122}
\affiliation{%
  \institution{Fudan University}
  \city{Shanghai}
  \country{China}
}
\email{cfsha@fudan.edu.cn}

\author{Xin Peng}
\orcid{0000-0003-3376-2581}
\affiliation{%
  \institution{Fudan University}
  \city{Shanghai}
  \country{China}
}
\email{pengxin@fudan.edu.cn}

%%
%% By default, the full list of authors will be used in the page
%% headers. Often, this list is too long, and will overlap
%% other information printed in the page headers. This command allows
%% the author to define a more concise list
%% of authors' names for this purpose.
\renewcommand{\shortauthors}{X. Zhou, Z. Dong, M. Ren, Q. Li, J. Li, S. Wang, X. Yu, C. Sha, and X. Peng}

%%
%% The abstract is a short summary of the work to be presented in the
%% article.
\begin{abstract}
Modern REST API testing faces a critical challenge in defining reliable test oracles, particularly in agile industrial environments where formal specifications (e.g., OpenAPI) are frequently missing or outdated, and historical execution logs are unavailable for newly deployed endpoints. In this paper, we present \toolname (\fulltoolname), a framework that generates executable test assertions from a single observed request--response pair in a black-box setting. Unlike existing approaches that rely on rule-based templates or massive training logs, \toolname utilizes a novel data augmentation pipeline to fine-tune a lightweight Large Language Model (LLM) via Group Relative Policy Optimization (GRPO). This training process enables the model to internalize testing ``common sense'' by optimizing a reward function that jointly encourages: (i) the selection of stable, semantically meaningful fields for validation and the avoidance of dynamic noise (e.g., timestamps or trace IDs); (ii) the generation of robust assertions that withstand logic variations. We evaluate \toolname on an industrial dataset comprising over 2,300 API traces across 246 real-world services. Comprehensive experiments demonstrate that \toolname significantly outperforms prompt-engineered baselines and generalist models, achieving a superior $F_1$~score of 85.42\% in key field identification and increasing the proportion of semantically accurate assertions. Furthermore, deployment in a production CI/CD workflow at ByteDance confirms its practical value: the system raised the adoption rate of automatically generated test cases from 74.1\% to over 96\%, substantially reducing manual Quality Assurance (QA) effort while ensuring high execution stability.
\end{abstract}

%%
%% The code below is generated by the tool at http://dl.acm.org/ccs.cfm.
%%
\begin{CCSXML}
<ccs2012>
   <concept>
       <concept_id>10011007.10011074.10011099.10011102.10011103</concept_id>
       <concept_desc>Software and its engineering~Software testing and debugging</concept_desc>
       <concept_significance>500</concept_significance>
       </concept>
   <concept>
       <concept_id>10002951.10003260.10003304.10003306</concept_id>
       <concept_desc>Information systems~RESTful web services</concept_desc>
       <concept_significance>500</concept_significance>
       </concept>
 </ccs2012>
\end{CCSXML}

\ccsdesc[500]{Software and its engineering~Software testing and debugging}
\ccsdesc[500]{Information systems~RESTful web services}

%%
%% Keywords. The author(s) should pick words that accurately describe
%% the work being presented. Separate the keywords with commas.
\keywords{REST API Testing, Oracle Generation, Deep Reinforcement Learning}

%%
%% This command processes the author and affiliation and title
%% information and builds the first part of the formatted document.
\maketitle

\section{Introduction}
\label{sec:intro}

In modern software architectures, REpresentational State Transfer (REST) has established itself as the \textit{de facto} standard for designing networked applications and microservices~\cite{fielding2000architectural,zhang2024trace,chen2023dynamic}. These APIs facilitate communication between heterogeneous components via standard HTTP methods and lightweight formats such as JSON. As systems evolve toward complex, distributed ecosystems, the reliability of these interfaces becomes paramount. A single regression in a backend API can propagate cascading failures to front-end applications, resulting in critical service disruptions or revenue leakage. Consequently, automated API testing has become an indispensable component of Continuous Integration/Continuous Deployment (CI/CD) pipelines, aimed at detecting faults efficiently before system release.

However, the rapid iteration cycles characteristic of agile development impose severe constraints on Quality Assurance (QA). In many industrial settings, particularly within fast-paced domains like video editing or content delivery, developers frequently deploy new features or modify existing logic without updating formal documentation. This results in missing, incomplete, or outdated specifications (e.g., OpenAPI/Swagger~\cite{openapi2025}). Furthermore, for newly deployed interfaces, massive historical execution logs are often unavailable, rendering statistical analysis impossible. This creates a ``cold-start'' testing scenario: QA engineers sometimes have to write robust test oracles based solely on a single traffic sample (one request-response pair) captured during a manual smoke test, relying on domain knowledge to infer assertion logic.

Existing automated testing approaches fail to adequately address this zero-specification, single-sample constraint. Traditional test oracle generation tools~\cite{alonso2023agora, alonso2025satori} typically depend on accurate specifications to generate schema-valid assertions; when the specification is absent, these tools are rendered ineffective. Similarly, dynamic invariant detection techniques (e.g., AGORA+~\cite{alonso2024agoraplus}) require large-scale historical logs to achieve statistical significance. Applying them to a single sample often yields fragile or trivial assertions that fail to capture complex business logic.

Recently, Large Language Models (LLMs) have shown promise in software testing tasks~\cite{schafer2023empirical,li2025llm,you2026industrial}. Yet, deploying general-purpose Large LLMs (e.g., DeepSeek-V3~\cite{deepseekai2024deepseekv3technicalreport}) for high-frequency API testing presents significant hurdles. First, the inference latency and financial cost of large parameter models are prohibitive for industrial CI/CD pipelines that execute thousands of tests daily. Second, without specific fine-tuning, general-purpose models often lack the necessary domain rigor; they tend to hallucinate testing logic or over-assert on fields unrelated to business logic, generating ``flaky'' tests that require constant maintenance. Using smaller, generic models alleviates cost but typically lacks the reasoning capability to generate assertions with high accuracy.

To bridge this gap, we present \toolname, a novel framework designed to generate actionable, industrial-grade test assertions from a single API traffic sample without reliance on formal specifications or historical logs. Unlike standard Supervised Fine-Tuning (SFT) approaches that require a massive ground-truth corpus of perfect code, our approach leverages Reinforcement Learning (RL). Specifically, we employ Group Relative Policy Optimization (GRPO)~\cite{shao2024deepseekmath} to fine-tune a lightweight LLM using an execution-feedback mechanism.

The core intuition behind \toolname is to model the ``common sense'' of a human tester. We construct a rewarding pipeline where the agent is optimized for semantic accuracy rather than mere syntax mimicry. By rewarding the agent for correctly handling augmented traffic variations, the model learns to independently distinguish valid data fluctuations from actual errors. This process enables the model to internalize the comprehensive boundaries of business logic and generate robust assertions, effectively bypassing the need for explicit rule definitions or extensive supervision.

We comprehensively evaluated \toolname through offline controlled experiments, expert user studies, and online production deployment at \companyname. On a curated test set of unseen API samples, \toolname achieved an $F_1$~score of 85.42\% in key field identification, outperforming the large-scale generalist model \texttt{DeepSeek-V3.1-Terminus}~\cite{deepseek2025terminus} by offering a superior balance of precision and recall. Regarding semantic correctness, \toolname generated the highest number of \textit{Exact Match} assertions (663) and minimized missed validations to only 28 cases, effectively solving the coverage issues observed in the base model. In qualitative user studies, domain experts confirmed that \toolname produced more actionable assertions with significantly reduced noise, requiring fewer manual edits than baseline approaches. Most notably, in the real-world production environment, the deployment of \toolname drove the Adoption Rate of generated test cases from 74.1\% to a sustained level above \textbf{96\%}, confirming its capability to meet strict industrial reliability standards.

The contributions of this paper are summarized as follows:

\begin{itemize}
    \item We propose a Reinforcement Learning-based framework, \toolname, that utilizes GRPO and a novel reward strategy to fine-tune a lightweight LLM. This allows the model to learn rigorous testing logic without requiring a ground-truth corpus of code.
    \item We provide a comprehensive evaluation and report on the large-scale industrial deployment of \toolname at \companyname. To our knowledge, this is the first study to demonstrate the practical efficacy of RL-fine-tuned LLMs for API oracle generation in a production CI/CD environment.
\end{itemize}

The remainder of this paper is organized as follows: Section~\ref{sec:background} provides the background and a motivating example. Section~\ref{sec:approach} details the \toolname approach, including dataset construction and model training. Section~\ref{sec:implementation} describes the implementation details. Section~\ref{sec:evaluation} presents a comprehensive evaluation, including offline experiments, the qualitative user study, and industrial deployment results. Section~\ref{sec:related_work} discusses related work. Section~\ref{sec:threats} outlines threats to validity, and Section~\ref{sec:conclusion} concludes the paper.

\section{Background and Motivating Example}\label{sec:background}

Consider the API \inlinecode{POST /api/subscription/plan\_list}, designed to retrieve user subscription details. Figure~\ref{fig:motivation_example} (left) presents a typical response body containing nested business objects. The response explicitly details the user's status through fields such as \inlinecode{cloud\_info} (cloud storage plans) and \inlinecode{vip\_info} (premium membership). Key attributes include \inlinecode{is\_using}, which indicates whether a specific plan is active; \inlinecode{subscribe\_type}, which defines the renewal policy (e.g., ``un-auto'' for non-renewing or ``auto'' for recurring billing); and temporal constraints defined by \inlinecode{begin\_time} and \inlinecode{end\_time}.

\begin{figure*}[t]
    \centering
    \lstset{
        basicstyle=\ttfamily\scriptsize, % Scriptsize fits columns better
        breaklines=true,
        keywordstyle=\color{blue},
        stringstyle=\color{green!50!black},
        columns=fullflexible,
        keepspaces=true,
        breakatwhitespace=false,
        numbers=left,
        numbersep=3pt,
        numberstyle=\tiny\color{gray},
        frame=single,
        xleftmargin=10pt,
        aboveskip=0pt,
        belowskip=0pt
    }

    % --- Caption Setup for this Figure ---
    % justification=raggedright: Prevents ugly stretching in narrow columns
    % singlelinecheck=false: Ensures even single lines obey the alignment
    % labelfont=bf: Makes "Listing 1" bold (optional, looks professional)
    \captionsetup[lstlisting]{
        justification=raggedright, 
        singlelinecheck=false, 
        font={scriptsize} 
    }

    \Description{A two-column figure showing a JSON API response on the left and its corresponding Python test assertions on the right. The JSON includes subscription plan details and timestamps. The Python code demonstrates logic for validating business constraints like start times being before end times.}

    % --- Left Column: JSON Response ---
    % \begin{minipage}[t]{0.48\textwidth}
    \begin{minipage}[t]{0.48\linewidth}
\begin{lstlisting}[
            language=Java, % logic works well for JSON highlighting
            % caption={Response example of \inlinecode{POST /subscription/plan\_list}.},
            label={lst:bg_response}
        ]
{
 "ret": "0",
 "errmsg": "success",
 "systime": "1751945691302",
 "log_id": "...",
 "data": {
  "cloud_info": {
   "subscribe_type": "un-auto",
   "plans": [
    {
     "package_name": "1024GB/Monthly Subscription",
     "begin_time": 1750832254,
     "end_time": 1753424254,
     "product_id": "pro_monthly",
     "is_using": true
    }
   ]
  },
  "vip_info": {
   "subscribe_type": "auto",
   "plans": [...]
  },
  ...
 }
}
\end{lstlisting}
    \end{minipage}
    \hfill
    % --- Right Column: Python Test ---
    % \begin{minipage}[t]{0.48\textwidth}
    \begin{minipage}[t]{0.48\linewidth}
\begin{lstlisting}[
            language=Python,
            % caption={Corresponding test case relying on domain "common sense".},
            label={lst:bg_assertion}
        ]
... # Test prefix

# Assume the response body is `resp`
# API status assertions
assert resp["ret"] == "0"
assert resp["errmsg"] == "success"
assert "cloud_info" in resp["data"]
cloud_info = resp["data"]["cloud_info"]
# Common sense: validating enum constraints
assert cloud_info["subscribe_type"] in ["auto", "un-auto"]

for plan in cloud_info["plans"]:
  # Common sense: validating temporal logic
  assert 0 < plan["begin_time"] < plan["end_time"]
  assert len(plan["package_name"]) > 0
  assert len(plan["product_id"]) > 0
  assert isinstance(plan["is_using"], bool)

# Similar assertions for `vip_info`
assert "vip_info" in resp["data"]
vip_info = resp["data"]["vip_info"]

for plan in vip_info["plans"]:
  ...



\end{lstlisting}
    \end{minipage}

    \vspace{-0.1cm}
    \caption{Motivating Example. Left: Response example of \inlinecode{POST /subscription/plan\_list}. Right: A manually written test case where QA engineers typically infer implicit constraints (e.g., temporal order or enumerations) relying on domain "common sense".}
    \label{fig:motivation_example}
\end{figure*}

In a rapid iterative development environment, server-side code modifications occur frequently, introducing the risk of unintended side effects or regressions. For instance, if \inlinecode{subscribe\_type} returns an unrecognized value (e.g., a typo or NULL), the billing system may fail to execute the renewal, directly leading to revenue leakage. Conversely, erroneous formatting could disrupt the front-end display. Consequently, Quality Assurance (QA) engineers are tasked with maintaining a robust suite of test cases to verify that the API functions correctly and that the data integrity involves strict semantic compliance.

To ensure reliability, test cases must go beyond simple HTTP status checks. Figure~\ref{fig:motivation_example} (right) illustrates a comprehensive test script for the aforementioned API. The assertions verify not only the presence of the \inlinecode{cloud\_info} structure (Line 7) but also enforce specific business logic. Crucially, writing these assertions often relies on the QA engineer's ``common sense'' and domain knowledge. For example, an engineer can infer that \inlinecode{subscribe\_type} should belong to a closed set \inlinecode{["auto", "un-auto"]} (Line 10) by observing different parts of the response, and intuitively asserts that a start timestamp should strictly precede an end timestamp (Line 14), even without explicit documentation.

However, manually writing such detailed assertions is labor-intensive and expensive, particularly given the scale of thousands of interfaces, each often containing numerous fields. While automated testing solutions exist, they face significant challenges in our industrial setting. Traditional schema-based tools require formal specifications (e.g., OpenAPI/Swagger), which are often absent or outdated in agile environments. Statistical approaches require massive historical traffic logs to infer patterns. In our scenario, we face a ``cold start'' problem: the testing platform operates in a black-box manner, relying on a single captured traffic sample (one request-response pair) without access to source code or database schemas.

The heavy reliance on ``common sense'' for generating meaningful assertions suggests that Large Language Models (LLMs) could be a viable solution, given their semantic reasoning capabilities. Nevertheless, deploying general-purpose Large LLMs (e.g., DeepSeek V3) within a high-frequency CI/CD pipeline is impractical due to high deployment costs and latency. Conversely, smaller, general-purpose models often lack the specific reasoning capabilities required to generate accurate assertions from limited context. To address this trade-off, we propose fine-tuning a lightweight model. By training a smaller model to internalize this testing-specific ``common sense,'' we aim to achieve high-quality assertion generation that satisfies both accuracy requirements and operational constraints (cost and speed).

\section{Approach} \label{sec:approach}

We present \toolname, a framework based on Reinforcement Learning (RL) designed to autonomously generate executable test assertions from isolated API traffic samples, even in the absence of explicit specifications (e.g., OpenAPI) or historical execution logs.

\begin{figure}[h]
    \centering
    \includegraphics[width=0.95\linewidth]{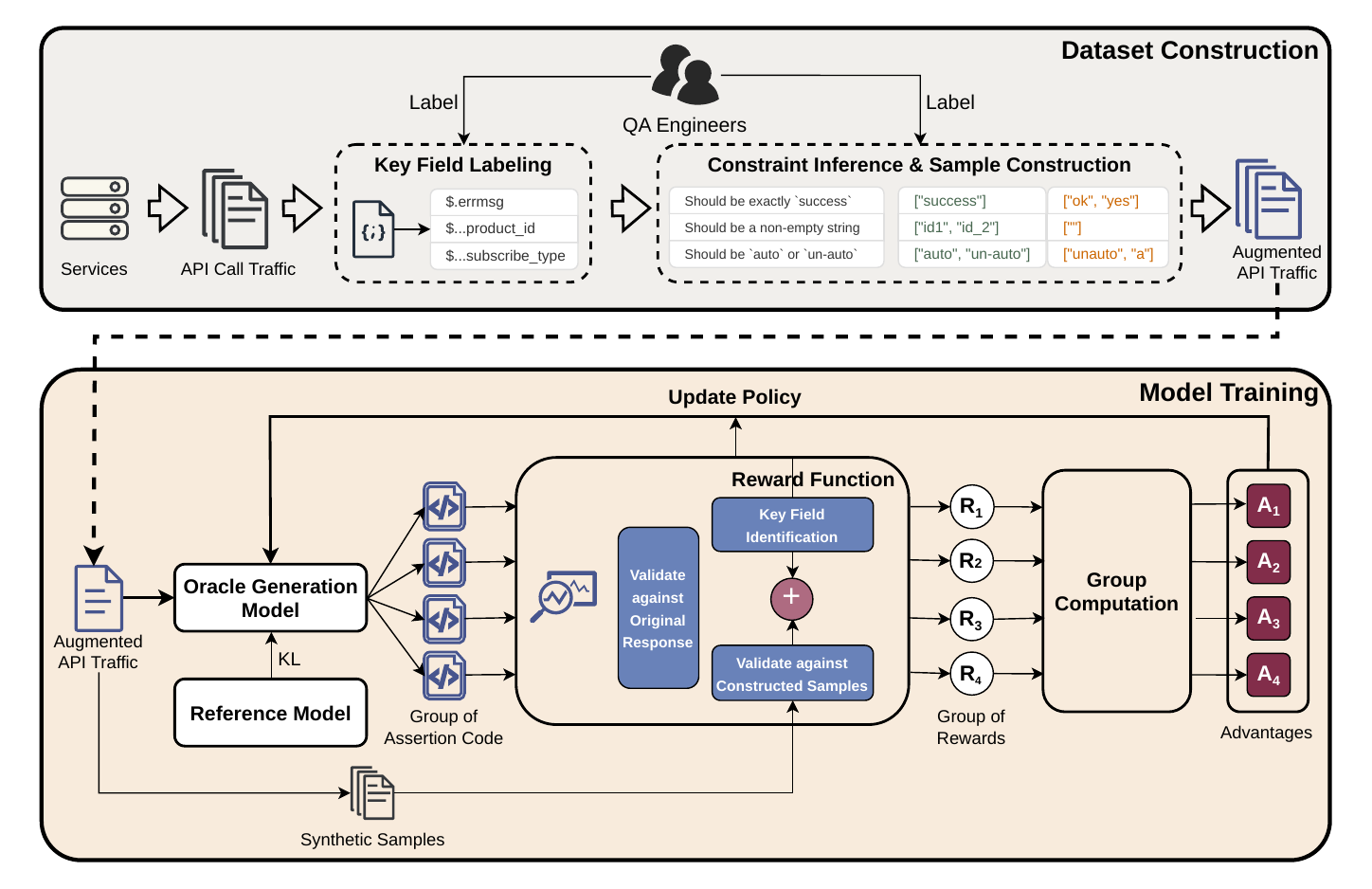}
    \caption{Overview of the \toolname framework. The workflow comprises two primary phases: (1) \textit{Dataset Construction} (top) and (2) \textit{Model Training} (bottom).}
    \Description{A diagram illustrating the two-phase process of \toolname: Data Augmentation extracting fields and constraints, followed by the RL-based training phase using GRPO.} 
    \label{fig:workflow_overview}
\end{figure}

As illustrated in Figure~\ref{fig:workflow_overview}, \toolname addresses the challenge of creating robust oracles from limited context through two coordinated phases. We first facilitate a data augmentation pipeline designed to enrich raw traffic by identifying key business fields and generating positive and negative samples based on inferred constraints.

Subsequently, we employ an RL-based training process where a Large Language Model (LLM) functions as the policy network. We adopt the Reinforcement Learning paradigm rather than traditional Supervised Fine-Tuning (SFT) for two primary reasons. First, test assertions lack a unique, canonical ``ground truth''; the same semantic constraint can be validated through diverse syntactic structures, making token-level imitation inefficient. Second, implicit field constraints are difficult to annotate precisely in a static dataset. RL allows the model to explore the solution space effectively, optimizing for functional correctness and logic coverage rather than strict syntactic matching.

Specifically, we utilize \textbf{Group Relative Policy Optimization (GRPO)}~\cite{shao2024deepseekmath} to fine-tune the model. As a state-of-the-art on-policy algorithm, GRPO optimizes the policy using group-based relative rewards without the need for an additional value network. This approach significantly reduces the memory footprint and computational overhead compared to traditional Actor-Critic methods (e.g., PPO), thereby providing a cost-effective solution for industrial-scale reasoning tasks while maintaining high generation performance.

\subsection{Dataset Construction} \label{sec:dataset}

To facilitate the training of the agent, we construct an augmented dataset $\mathcal{D}_{\text{aug}}$ derived from real-world production traffic. Each training instance $T^{\text{aug}} \in \mathcal{D}_{\text{aug}}$ is represented as a tuple that encapsulates not only the raw API call traffic but also the semantic metadata required for rigorous assessment:
\[
    T^{\text{aug}} = \langle \text{API}, \text{req}, \text{resp}, \mathbb{K}, \mathcal{S}_{\text{pos}}, \mathcal{S}_{\text{neg}} \rangle
\]
where:
\begin{itemize}
    \item $\langle \text{API}, \text{req}, \text{resp} \rangle$ represents the original captured traffic, containing the endpoint identifier, request payload, and the raw server response.
    \item $\mathbb{K}$ denotes the set of \textit{Key Fields} extracted from the response, which serve as the focal point for assertion generation.
    \item $\mathcal{S}_{\text{pos}}$ and $\mathcal{S}_{\text{neg}}$ are sets of generated response bodies representing valid (positive) and invalid (negative) variations of the original data, respectively.
\end{itemize}

The construction of $T^{\text{aug}}$ involves two primary phases: identifying which fields require testing ($\mathbb{K}$) and generating data variations ($\mathcal{S}_{\text{pos}}, \mathcal{S}_{\text{neg}}$) to evaluate the quality of the generated assertions.

\subsubsection{Collecting Traffic Data}
Initially, we collect production traffic records from an internal API recording platform. Each record, denoted as $T$, encapsulates a single API interaction, comprising the API endpoint identifier ($\text{API}$), the client request payload ($\text{req}$), and the server response body ($\text{resp}$):
\[
    T = \langle \text{API}, \text{req}, \text{resp} \rangle
\]
\toolname primarily focuses on analyzing the $\text{resp}$ component, which is typically a semi-structured document (e.g., JSON) representing a hierarchical collection of field-value pairs.

\subsubsection{Key Field Labeling}
Raw API responses frequently contain noise, such as transient diagnostic data or tracing identifiers, which are irrelevant to functional correctness. To ensure that the model focuses its attention on business-critical logic rather than structural noise, we filter the response fields to obtain the set $\mathbb{K}$.

To guarantee the reliability and precision of the dataset, the labeling process was performed through manual annotation by three professional QA engineers from \companyname. To mitigate subjective bias, a majority voting mechanism was implemented to determine the final composition of $\mathbb{K}$. Specifically, each field was independently reviewed by the experts, and only those selected by at least two out of the three annotators were retained. Guided by deep domain expertise, this selection process adheres to three governing principles:

\begin{enumerate}
    \item \textbf{Entity Identification:} Fields that uniquely identify business entities (e.g., \inlinecode{product\_id}) are included to ensure data integrity.
    \item \textbf{Domain Logic Constraints:} Experts label fields conveying essential domain logic based on their experience. For example, time intervals (e.g., \inlinecode{begin\_time}, \inlinecode{end\_time}) and some enumeration fields (e.g., \inlinecode{subscribe\_type}), are selected to verify semantic compliance.
    \item \textbf{Operational Status:} Fields indicating the outcome of the request (e.g., \inlinecode{ret} codes, \inlinecode{errmsg}) are retained to validate the execution state.
\end{enumerate}
Conversely, dynamic fields such as \inlinecode{log\_id} or server timestamps (e.g., \inlinecode{systime}) are explicitly excluded to maintain deterministic testing conditions.

\subsubsection{Semantic Constraint Inference and Evaluation Sample Construction}
Since the dataset lacks formal specifications (e.g., OpenAPI), we cannot directly determine whether a generated assertion is factually correct. To address this, we construct positive and negative samples to serve as a testbed for evaluating the quality of model-generated assertions. The underlying premise is that a high-quality assertion should pass validation against all positive samples ($\mathcal{S}_{\text{pos}}$) while effectively flagging all negative samples ($\mathcal{S}_{\text{neg}}$).

To ensure the accuracy of the inferred semantic constraints and the representativeness of the constructed data, we employed the same rigorous annotation protocol described in the previous section. Specifically, the three expert QA engineers independently formulated the semantic constraints and designed the corresponding sample variations, with final decisions determined via a majority voting mechanism. For each field in $\mathbb{K}$, this process yields an enriched Natural Language (NL) constraint based on the field's name and its value in the original response. For example, as shown in Table~\ref{tab:field_constraints}, the confirmed constraint for \inlinecode{begin\_time} is ``Should be a non-negative Unix timestamp.'' Guided by these verified NL constraints, we then construct two sets of samples:

\begin{itemize}
    \item \textbf{Positive Samples ($\mathcal{S}_{\text{pos}}$):} We construct response bodies where the values of key fields are replaced but remain compliant with the inferred constraints. For instance, replacing the timestamp \texttt{1750832254} with \texttt{1750832255} creates a valid variation. These samples ensure the generated assertions are not overfitting to the specific values of the single captured traffic.
    \item \textbf{Negative Samples ($\mathcal{S}_{\text{neg}}$):} We construct response bodies containing deliberate violations of the constraints. For example, injecting \texttt{-1} into a timestamp field or a string into a boolean field (see Table~\ref{tab:field_constraints}). These samples are crucial for verifying that the generated assertions possess sufficient strictness to detect data anomalies.
\end{itemize}

For each identified key field, we replace the original field value with three valid ones and three invalid ones to construct response variants. Some fields such as status code may have only one correct value, and we do not construct valid samples for them.

\begin{table}[h]
    \caption{Augmentation Examples for Key Fields in \inlinecode{POST /api/subscription/plan\_list}}
    \label{tab:field_constraints}
    \centering
    \small
    \begin{tabular}{ >{\raggedright\arraybackslash}p{8em} | >{\raggedright\arraybackslash}p{10em} | >{\raggedright\arraybackslash}p{8em} | >{\raggedright\arraybackslash}p{8em} }        \toprule
        \textbf{Field} & \textbf{Inferred Constraint} & \textbf{Constructed Positive Value} & \textbf{Constructed Negative Value} \\
        \midrule
        \inlinecode{ret} & Should be '0' for success. & \texttt{"0"} & \texttt{"1"} \\
        \midrule
        \inlinecode{errmsg} & Should be 'success'. & \texttt{"success"} & \texttt{"internal error"} \\
        \midrule
        \inlinecode{...product\_id} & Should be a non-empty string. & \texttt{"another\_id"} & \texttt{""} \\
        \midrule
        \inlinecode{...begin\_time} & Should be a non-negative Unix timestamp. & \texttt{1750832255} & \texttt{-1} \\
        \midrule
        \inlinecode{...subscribe\_type} & Should be one of "auto" or "un-auto". & \texttt{"auto"} & \texttt{"manual"} \\
        \midrule
        \inlinecode{...is\_using} & Should be a boolean value. & \texttt{false} & \texttt{"true"} \\
        \bottomrule
    \end{tabular}
\end{table}

\subsection{Model Training}

\subsubsection{Oracle Generation via Group Relative Policy Optimization}

We formulate oracle generation as a reinforcement learning task where the
policy $\pi_\theta$ generates assertion code conditioned on the API context.
To efficiently align the model with test correctness objectives without the
computational cost of training a value function, we employ \textbf{Group
Relative Policy Optimization (GRPO)}~\cite{shao2024deepseekmath}.

For an API context $q$ sampled from distribution $\mu$, GRPO samples a group
of outputs $\{o_1,o_2,\ldots,o_G\}$ from the old policy
$\pi_{\theta_{\mathrm{old}}}(\cdot\mid q)$. Let
$o_i=(o_{i,1},\ldots,o_{i,|o_i|})$ denote the token sequence of the $i$-th
output. The policy is optimized by maximizing the following objective:
\begin{equation}
\begin{split}
\mathcal{J}_{\mathrm{GRPO}}(\theta)
={}&
\mathbb{E}_{\substack{
q\sim\mu,\,
\{o_i\}_{i=1}^{G}\sim
\pi_{\theta_{\mathrm{old}}}(\cdot\mid q)
}}
\Bigg[
\frac{1}{G}\sum_{i=1}^{G}\frac{1}{|o_i|}
\sum_{t=1}^{|o_i|}
\Bigg\{
\\
&\min\left[
\frac{
\pi_\theta(o_{i,t}\mid q,o_{i,<t})
}{
\pi_{\theta_{\mathrm{old}}}(o_{i,t}\mid q,o_{i,<t})
}
\hat{A}_{i,t},
\right.
\\[-0.1cm]
&\left.
\operatorname{clip}\left(
\frac{
\pi_\theta(o_{i,t}\mid q,o_{i,<t})
}{
\pi_{\theta_{\mathrm{old}}}(o_{i,t}\mid q,o_{i,<t})
},
1-\epsilon,
1+\epsilon
\right)
\hat{A}_{i,t}
\right]
-
\beta\mathbb{D}_{KL}
\left[\pi_\theta\|\pi_{\mathrm{ref}}\right]
\Bigg\}
\Bigg].
\end{split}
\label{equ:Jgrpo}
\end{equation}
Here, $\epsilon$ is the clipping hyperparameter, $\beta$ controls the
KL-divergence penalty, and $\hat{A}_{i,t}$ is the advantage assigned to the
$t$-th token of output $o_i$. Unlike PPO, GRPO estimates the advantage from
the relative rewards of outputs within the same group and therefore does not
require a separate value model.

Following GRPO, the token-level KL divergence between the current policy and
the reference policy is estimated as
\begin{equation}
\mathbb{D}_{KL}\left[\pi_\theta\|\pi_{\mathrm{ref}}\right]
=
\frac{
\pi_{\mathrm{ref}}(o_{i,t}\mid q,o_{i,<t})
}{
\pi_\theta(o_{i,t}\mid q,o_{i,<t})
}
-
\log
\frac{
\pi_{\mathrm{ref}}(o_{i,t}\mid q,o_{i,<t})
}{
\pi_\theta(o_{i,t}\mid q,o_{i,<t})
}
-1.
\end{equation}

In our setting, each generated assertion $o_i$ receives an outcome-level
reward $r_i$. The rewards within a group are normalized by subtracting their
mean and dividing by their standard deviation. The resulting normalized
reward is assigned as the advantage of every token in the corresponding
output:
\begin{equation}
\hat{A}_{i,t}
=
\frac{
r_i-\operatorname{mean}(r_1,\ldots,r_G)
}{
\operatorname{std}(r_1,\ldots,r_G)
},
\qquad
t=1,\ldots,|o_i|.
\label{equ:ai}
\end{equation}
This group-relative normalization serves as a baseline and captures the
relative quality of each generated assertion without relying on a separate
critic network.

\paragraph{Reward Definitions}
The scalar reward $r_i$ in Equation~(\ref{equ:ai}) evaluates the assertion's
executability, key-field coverage, and semantic accuracy. Validating these
properties requires interacting with the environment's augmented data
samples. We detail the specific reward design in the following part.

\subsubsection{Reward Function}
\label{subsubsec:reward}

To guide the reinforcement learning agent toward generating test oracles that are syntactically correct, semantically precise, and functionally robust, we design a composite reward mechanism. The final reward $\mathcal{R}$ for a generated assertion $\mathcal{A}$ is formulated as a conditional function that first penalizes execution failures and then linearly combines field identification and semantic accuracy, clipped to a normalized range:

\begin{equation}
    \mathcal{R}(\mathcal{A}) = 
    \begin{cases} 
        \rho_{\text{fail}} & \text{if } \text{Exec}(\mathcal{A}, \text{resp}) = \text{False} \\
        \text{clip}\left( \alpha \cdot \mathcal{R}_{\text{ident}}(\mathcal{A}) + \beta \cdot \mathcal{R}_{\text{sem}}(\mathcal{A}), -1, 1 \right) & \text{otherwise}
    \end{cases}
    \label{eq:total_reward}
\end{equation}

where $\text{Exec}(\mathcal{A}, \text{resp})$ denotes the execution result on the original ground-truth response (returning \text{True} only if the assertion passes without error), and $\rho_{\text{fail}}$ is a fixed penalty (set to $-1$) for invalid assertions. For executable assertions, the reward is derived from two sub-components: Key Field Identification ($\mathcal{R}_{\text{ident}}$) and Semantic Accuracy ($\mathcal{R}_{\text{sem}}$), balanced by coefficients $\alpha$ and $\beta$. The final score is clipped to the interval $[-1, 1]$ to ensure optimization stability during the GRPO training phase. The components are defined as follows:

\paragraph{1. Validity Check (The Gatekeeper)}
The most fundamental requirement is that the generated assertion must validate the original captured API response successfully. If the assertion $\mathcal{A}$ raises an exception or asserts false on the observed data $\text{resp}$, it is fundamentally flawed. In such cases, the evaluation terminates immediately with the penalty $\rho_{\text{fail}}$, discouraging the model from generating hallucinatory or syntactically invalid code.

\paragraph{2. Key Field Identification ($\mathcal{R}_{\text{ident}}$)}
To ensure the assertion focuses on business-critical logic rather than irrelevant noise (e.g., log IDs), we calculate the overlap between the fields accessed by the assertion code, denoted as $\mathcal{F}(\mathcal{A})$, and the pre-identified key field set $\mathbb{K}$. We quantify relevance using precision and recall:
\[
    \text{precision} = \frac{|\mathcal{F}(\mathcal{A}) \cap \mathbb{K}|}{|\mathcal{F}(\mathcal{A})|}, \quad
    \text{recall} = \frac{|\mathcal{F}(\mathcal{A}) \cap \mathbb{K}|}{|\mathbb{K}|}
\]
The relevance reward is a weighted sum of precision and recall:
\[
    R_{\text{ident}}(\mathcal{A}) = w_{\text{prec}} \cdot \text{precision} + w_{\text{rec}} \cdot \text{recall}
\]
where $w_{\text{prec}}$ and $w_{\text{rec}}$ balance the trade-off between avoiding noise and ensuring comprehensive coverage.

\paragraph{3. Semantic Accuracy ($\mathcal{R}_{\text{sem}}$)}
A high-quality oracle must not only accept the observed traffic but also correctly distinguish between valid and invalid data distributions. We evaluate this using the augmented datasets generated in Section~\ref{sec:dataset}: the positive samples $\mathcal{S}_{\text{pos}}$ (valid variations) and negative samples $\mathcal{S}_{\text{neg}}$ (constraint violations).

Let $KillRate(\mathcal{A}, \mathcal{S}) \in [0, 1]$ represent the \textbf{Kill Rate} of assertion $\mathcal{A}$ over a set of samples $\mathcal{S}$, defined as the proportion of samples for which the assertion evaluates to false. This reward is defined as the differential behavior between these two sets:
\begin{equation}
    \mathcal{R}_{\text{sem}}(\mathcal{A}) = w_{\text{neg}} \cdot KillRate(\mathcal{A}, \mathcal{S}_{\text{neg}}) - w_{\text{pos}} \cdot KillRate(\mathcal{A}, \mathcal{S}_{\text{pos}})
\end{equation}
where $w_{\text{neg}}$ and $w_{\text{pos}}$ are hyperparameters.

Intuitively, this function rewards the model for maximizing the acceptance of valid positive samples (avoiding false positives/overfitting) while simultaneously maximizing the rejection of invalid negative samples (ensuring strictness/detection capability). An ideal assertion yields a pass rate of $1.0$ for $\mathcal{S}_{\text{pos}}$ and $0.0$ for $\mathcal{S}_{\text{neg}}$, resulting in a maximal component reward.

\section{Implementation}\label{sec:implementation}

The policy network of \toolname, responsible for generating assertion code token-by-token, is instantiated using \texttt{\baseModelFull}~\cite{doubao2025flash}. This choice is strategic; \baseModelFull is a lightweight and fast large language model developed internally at \companyname, enabling efficient iteration during training and deployment. To optimize the policy network, we employ GRPO~\cite{shao2024deepseekmath} as the fine-tuning algorithm, directly maximizing the expected reward objectives defined in the previous section.

All model training and experimentation were conducted on the internal cloud AI service platform at \companyname, utilizing standard computational resources for large-scale model development. Regarding the hyperparameter settings for GRPO, the model was fine-tuned for 2 epochs with a learning rate of $1 \times 10^{-6}$. To ensure training stability, the KL divergence coefficient was set to 0.001, and the policy clip ratio was fixed at 0.2. During the rollout phase, the number of generations was set to 8, meaning the policy sampled 8 distinct outputs for each training prompt to calculate the group-relative advantage. Due to the substantial computational resources and high financial cost required for reinforcement learning fine-tuning on Large Language Models, the training and evaluation processes were conducted in a single run.

\section{Evaluation}\label{sec:evaluation}

To comprehensively evaluate the performance of \toolname, we investigated the following three research questions:

\begin{itemize}
    \item{\myrq{1}} (\textbf{Effectiveness}): How effective is \toolname in generating test oracles for REST APIs compared to baseline models?
    \item{\myrq{2}} (\textbf{Expert Evaluation of Usefulness and Reliability}): How do software testing experts evaluate usefulness, and reliability of the generated assertions?
    \item{\myrq{3}} (\textbf{Industrial Application}): How does \toolname perform when deployed in production environment at \companyname?
\end{itemize}

\subsection{Experimental Setup}
\label{subsec:exp_setup}

\subsubsection{Test Generation Platform}

\toolname is integrated into our internal automated test case generation platform deployed at \companyname. This platform facilitates a workflow: QA engineers upload anonymized API traffic captured from the production or test environment, and the system generates a complete Python test script containing assertions. Engineers then review the generated code, modifying or accepting it for commit to the version control system. This infrastructure serves as the deployment environment for assessing industrial applicability.

\subsubsection{Dataset Construction}

We constructed a large-scale dataset of production API traffic sourced from an internal recording platform at \companyname. The dataset comprises over 2,300 unique traffic samples derived from over 1,500 distinct REST APIs (with each API contributing 1 to 3 traces). This spans 246 distinct services and 15 business lines, ensuring a diverse representation of industrial scenarios ranging from content management to billing systems. To protect privacy and confidentiality, all recorded traffic was anonymized prior to analysis, removing or irreversibly transforming any personally identifiable information and other sensitive identifiers to prevent disclosure of individual users.

% The dataset was randomly partitioned into training, validation, and testing sets in an 8:1:1 ratio. The main evaluation (\myrq{1}) is conducted on the held-out test set, consisting of 229 unseen API samples. As detailed in Section~\ref{sec:approach}, ground truth was established through expert annotation, where key fields were explicitly labeled and logical constraints were described in natural language. We utilized a voting consensus mechanism among three QA engineers to validate the semantic alignment of the ground truth.

The dataset was randomly partitioned into training, validation, and testing sets in an 8:1:1 ratio. The main evaluation (\myrq{1}) is conducted on the held-out test set, consisting of 229 unseen API samples. As detailed in Section~\ref{sec:approach}, ground truth was established through expert annotation, where key fields were explicitly labeled and logical constraints were described in natural language. We utilized a voting consensus mechanism among three QA engineers to validate the semantic alignment of the ground truth. This manual annotation process required substantial effort: the identification of key fields took approximately 4 days, while the definition of semantic constraints and the construction of positive and negative samples required approximately 2 weeks of effort from the annotators.

\subsubsection{Baselines}
We compare \toolname against two baseline models that rely solely on prompt engineering without reinforcement learning fine-tuning. To ensure a fair comparison, all models utilize the exact same system prompt and input format, the full details of which are provided in Appendix~\ref{appendix:prompt}. Furthermore, both baseline models operate in standard generation mode with reasoning capabilities disabled.

\begin{itemize}
    \item \textbf{\baseModelFull:} We evaluate the original, pre-trained version of the base model. Since this model serves as the initialization for our policy network, comparing against it quantifies the specific performance gains attributable to the GRPO fine-tuning process.
    \item \textbf{DeepSeek-V3.1-Terminus~\cite{deepseek2025terminus}:} We include DeepSeek-V3.1-Terminus (hereafter referred to as DeepSeek), one of the best open-source LLMs, derived from the \texttt{DeepSeek-V3}~\cite{deepseekai2024deepseekv3technicalreport} architecture. As a large-parameter foundation model, this baseline represents the capabilities of a powerful, general-purpose LLM. This comparison evaluates the trade-offs between a specialized lightweight model and a large-scale generalist model.
\end{itemize}

\subsubsection{Evaluation Metrics}
We assess model performance across two primary dimensions:

\paragraph{\textbf{1) Key Field Identification}}
This metric measures the model's ability to identify business-critical fields ($\mathbb{K}$) while filtering out structural noise (e.g., dynamic timestamps, request IDs). We report \textit{precision}, \textit{recall}, and \textit{$F_1$~score}.

\paragraph{\textbf{2) Assertion Accuracy}}
We evaluate the semantic correctness of the generated assertions. Based on expert review, each generated assertion is categorized as:
\begin{itemize}
    \item \textit{Exact Match:} The logic is semantically correct and aligns with domain requirements.
    \item \textit{Overly General:} The assertion is valid but too loose (e.g., checking for non-null instead of a specific enum value).
    \item \textit{Partial Match:} The code covers some semantic conditions but misses boundary details.
    \item \textit{Incorrect:} The generated logic is wrong (false positive).
    \item \textit{Not Validated:} The model failed to generate an assertion for a requisite key field.
\end{itemize}

\paragraph{\textbf{3) Industrial Adoption Rate}}
To quantify the practical utility of \toolname in the production environment (\myrq{3}), we utilize the \textbf{Adoption Rate}, derived from the platform described in Section~\ref{subsec:exp_setup}. This metric is defined as the ratio of generated test cases that are explicitly accepted and committed by QA engineers to the repository versus the total number of generation tasks. A high adoption rate indicates that the generated assertions meet strict industrial standards with minimal need for manual modification.

\subsection{Analysis of RQ1: Effectiveness}

In this section, we analyze the quality of generated assertions on the curated test set, focusing on two key dimensions: the ability to identify business-critical fields and the semantic correctness of the generated logic.

\subsubsection{Key Field Identification}

\begin{figure*}[t]
    \centering
    \includegraphics[width=0.9\textwidth]{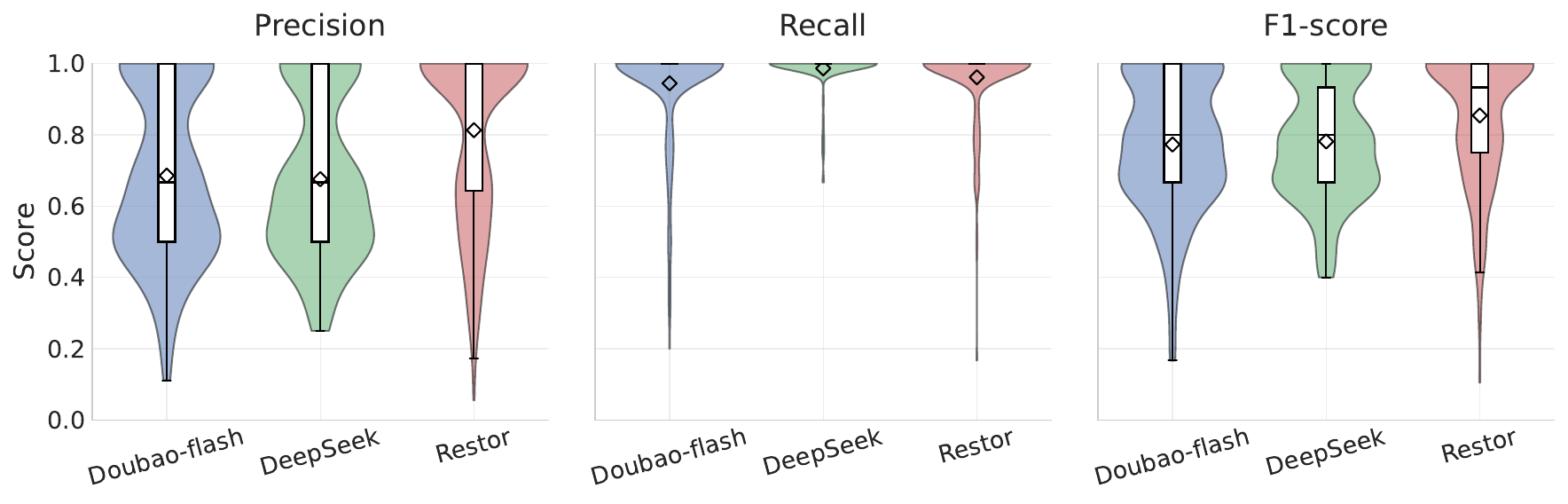}
    \Description{A set of three violin plots comparing precision, recall, and $F_1$~score across \baseModelShort, DeepSeek, and \toolname. Each violin plot contains a box plot indicating quartiles and a diamond marker indicating the mean value.}
    \caption{Distribution of Key Field Identification Performance. The visualization integrates violin plots to show data distribution density, box plots to indicate quartiles, and white diamond markers to denote the mean score.}
    \label{fig:rq1_field_identification}
\end{figure*}

We first evaluate the capability of the models to correctly identify the set of business-critical fields $\mathbb{K}$, distinguishing them from dynamic noise (e.g., \inlinecode{log\_id}, timestamps). Figure~\ref{fig:rq1_field_identification} visualizes the distribution of precision, recall, and $F_1$~score across the test set.

As indicated by the distributions, \toolname demonstrates a superior balance between coverage and selectivity. By leveraging GRPO to internalize the distinction between signal and noise, \toolname achieves a precision of 81.30\% and maintains a competitive recall of 96.15\%, consequently attaining the highest $F_1$~score of 85.42\%. The density distribution in Figure~\ref{fig:rq1_field_identification} corroborates this stability, as the $F_1$~score violin for \toolname is structurally tighter and concentrated at higher values compared to the dispersed distributions of the baseline models. 

In comparison, the baseline models exhibit distinct performance trade-offs. As indicated by the recall distribution, the large-scale model DeepSeek achieves the highest average recall of 98.64\%, reflecting its broad semantic coverage; however, its corresponding precision is notably limited at 67.57\%, indicating a tendency to over-generate assertions for irrelevant fields. Meanwhile, the base \baseModelShort model performs worse, exhibiting low precision (68.54\%) and a moderate $F_1$~score (77.29\%). 

To rigorously validate these comparisons, we conducted a one-tailed Wilcoxon signed-rank test ($N = 229$, confidence level 0.95). The statistical results confirm that \toolname achieves significant improvements in both precision and $F_1$~score when compared to \baseModelShort ($p < 0.001$) and DeepSeek ($p < 0.001$). While DeepSeek exhibits a slightly higher average recall, its significantly lower precision ultimately yields a statistically inferior $F_1$~score.

% For detailed results of Wilcoxon signed-rank test, see [file](codes/rq3_case_study_labels.json).

\subsubsection{Assertion Accuracy}

\begin{figure*}[t]
    \centering
    \includegraphics[width=0.9\textwidth]{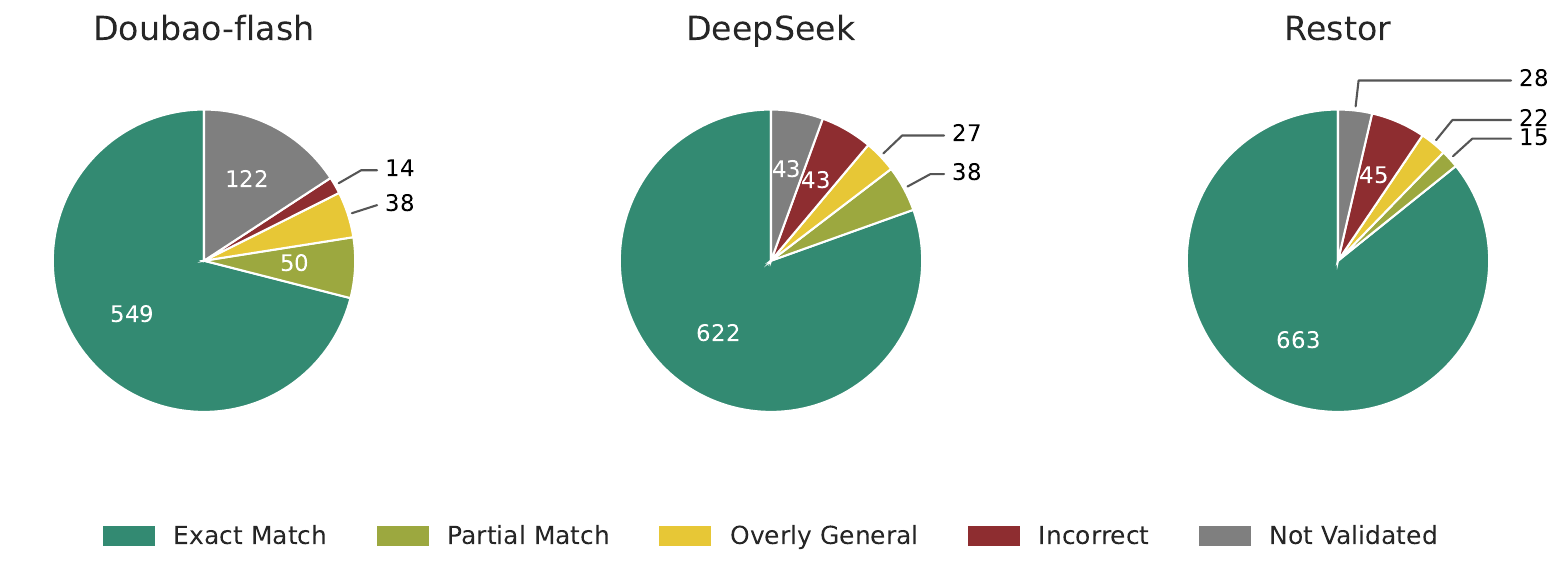}
    \Description{Three pie charts comparing the assertion accuracy distribution for \baseModelShort, DeepSeek, and \toolname. The charts show categories for Exact Match, Partial Match, Overly General, Incorrect, and Not Validated.}
    \caption{Comparative Analysis of Assertion Accuracy. This distribution quantifies the semantic correctness of the generated test code, highlighting the volume of exact matches versus missed validations.}
    \label{fig:rq1_assertion_accuracy}
\end{figure*}

Figure~\ref{fig:rq1_assertion_accuracy} presents the breakdown of assertion quality into five categories as defined in the experimental setup.

\toolname exhibits the highest efficacy, generating \textbf{663} \textit{Exact Match} assertions, surpassing both DeepSeek (622) and \baseModelShort (549). Critically, \toolname maximizes coverage completeness by minimizing the \textit{Not Validated} category. While \baseModelShort failed to validate 122 requisite fields and DeepSeek missed 43, \toolname reduced this figure to just \textbf{28}. This indicates that the RL fine-tuning process successfully aligned the output with the strict logical boundaries required for automated testing, reducing the need for manual correction.

\textbf{Handling False Positives and Non-determinism.} A critical factor in the practical usefulness of automated test generation is the rate of false positives (categorized here as \textit{Incorrect} assertions). Our evaluation confirms that such occurrences are exceptionally rare with \toolname. Furthermore, dynamic and non-deterministic fields (e.g., system timestamps or request IDs) are explicitly filtered during training and inference, preventing the generation of flaky assertions. In the production environment, any residual false positives are effectively mitigated through a ``human-in-the-loop'' mechanism: QA engineers review and refine the generated test cases prior to committing them to the code repository, ensuring that flawed assertions never disrupt automated CI/CD pipelines.

\subsubsection{Case Study}
\label{sec:case_study}

To qualitatively assess model performance, we analyze assertions generated for the API \inlinecode{POST /api/subscription/plan\_list}. Table~\ref{tab:rq1_assertion_comparison} presents a comparative analysis of how \toolname and the baselines handle specific critical fields and dynamic noise.

The baseline models exhibit distinct failure patterns. \baseModelShort predominantly relies on shallow structural checks, often validating only the existence of keys (e.g., \inlinecode{'quota' in space\_info}) rather than their values, which limits the depth of testing. Conversely, DeepSeek, creates two issues despite its reasoning capability: (1) Generic Constraints, where it defaults to weak checks (e.g., \inlinecode{len > 0}) instead of inferring Enums or value ranges, and (2) Noise Sensitivity, where it generates assertions for unstable system fields (e.g., \inlinecode{log\_id}), leading to flaky tests in production.

In contrast, \toolname demonstrates the ability to infer strict semantic boundaries. It correctly identifies enumeration constraints (e.g., \inlinecode{subscribe\_type}), enforces semantic type validation on string-encoded numbers, and successfully filters out volatile system fields, ensuring the assertions are both rigorous and stable.

\begin{table}[t]
    \caption{Comparison of Generated Assertions for Key Fields. For clarity, the \inlinecode{assert} keyword is omitted, and specific hierarchical field paths (e.g., \inlinecode{resp['data']['space\_info']['quota']}) are abstracted as \inlinecode{val}.}
    \label{tab:rq1_assertion_comparison}
    \centering
    \small
    \renewcommand{\arraystretch}{1.3}
    % 使用 >{\raggedright\arraybackslash} 强制每一列左对齐，消除单词间的大间距
    \begin{tabular}{
        >{\raggedright\arraybackslash}p{0.16\textwidth} | 
        >{\raggedright\arraybackslash}p{0.17\textwidth} | 
        >{\raggedright\arraybackslash}p{0.17\textwidth} | 
        >{\raggedright\arraybackslash}p{0.17\textwidth} | 
        >{\raggedright\arraybackslash}p{0.17\textwidth}
    }
        \toprule
        \textbf{Target Field} & \textbf{Ground Truth Constraint} & \textbf{\baseModelShort} & \textbf{DeepSeek} & \textbf{\toolname} \\
        \midrule

        \inlinecode{errmsg} & 
        Should be 'success'. & 
        \textbf{Exact Match:} \newline \inlinecode{val == 'success'} \newline  & 
        \textbf{Exact Match:} \newline \inlinecode{val == 'success'} & 
        \textbf{Exact Match:} \newline \inlinecode{val == 'success'} \\
        \midrule
        
        \inlinecode{subscribe\_type} & 
        Should be one of 'auto' or 'un-auto'. & 
        \textit{(Not Validated)} & 
        \textbf{Overly General:} \newline \inlinecode{len(val) > 0} & 
        \textbf{Exact Match:} \newline \inlinecode{val in ['un-auto', 'auto']} \\
        \midrule
        \inlinecode{space\_info.quota} & 
        Should be a string representing a non-negative integer. & 
        \textbf{Overly General:} \newline \inlinecode{'quota' in space\_info} & 
        \textbf{Partial Match:} \newline \inlinecode{int(val) >= 0} & 
        \textbf{Exact Match:} \newline \inlinecode{isinstance(str) and int(val) >= 0} \\
        \midrule
        \inlinecode{plan.begin\_time} & 
        Should be a non-negative Unix timestamp. & 
        \textbf{Overly General:} \newline \inlinecode{'begin\_time' in plan} & 
        \textbf{Partial Match:} \newline \inlinecode{int(val) >= 0} & 
        \textbf{Partial Match:} \newline \inlinecode{val >= 0} \\
        \midrule

        \inlinecode{systime} / \inlinecode{log\_id} & 
        \textit{\textbf{Noise:} Dynamic system fields; should be ignored.} & 
        \textit{(Ignored)} & 
        \textbf{False Positive:} \newline \inlinecode{int(val) > 0} \newline \inlinecode{len(val) > 0} & 
        \textit{(Ignored)} \\
        \bottomrule
    \end{tabular}
\end{table}

\begin{tcolorbox}[colback=LightGray, title=Answer to \myrq{1}, colframe=black, left=1mm, right=1mm, top=1mm, bottom=1mm]
\toolname significantly improves upon prompt-based baselines in automated assertion generation. It achieves the highest $F_1$~score (\textbf{85.42\%}) in identifying business-critical fields, effectively filtering dynamic noise that plagues large general-purpose models. Furthermore, \toolname produces the highest volume of \textit{Exact Match} assertions while minimizing missed validations, demonstrating that GRPO fine-tuning enables a lightweight model to generate stricter, more complete, and production-ready test oracles.
\end{tcolorbox}

% Labeled key fields and their constraints: see codes/rq3_case_study_labels.json

\subsection{Analysis of RQ2: Expert Evaluation of Usefulness and Reliability}
\label{sec:rq2}

This section evaluates the practical utility of \toolname in real-world testing workflows. We conducted a study on 48 cases randomly sampled from our internal automated test case generation platform. Users uploaded API traffic and used our platform to generate these cases, which exhibit higher structural variance and noise than the curated dataset. Different from the curated setting in \myrq{1}, we collect both (i) quantitative evaluation results and (ii) qualitative feedback from users, including their perceived key fields, preferred assertion strictness, and additional usability concerns such as generation latency, code style, and maintainability. The goal is to determine whether the generated oracles are useful for our users.

\begin{figure}[h]
    \centering
    \includegraphics[width=0.9\linewidth]{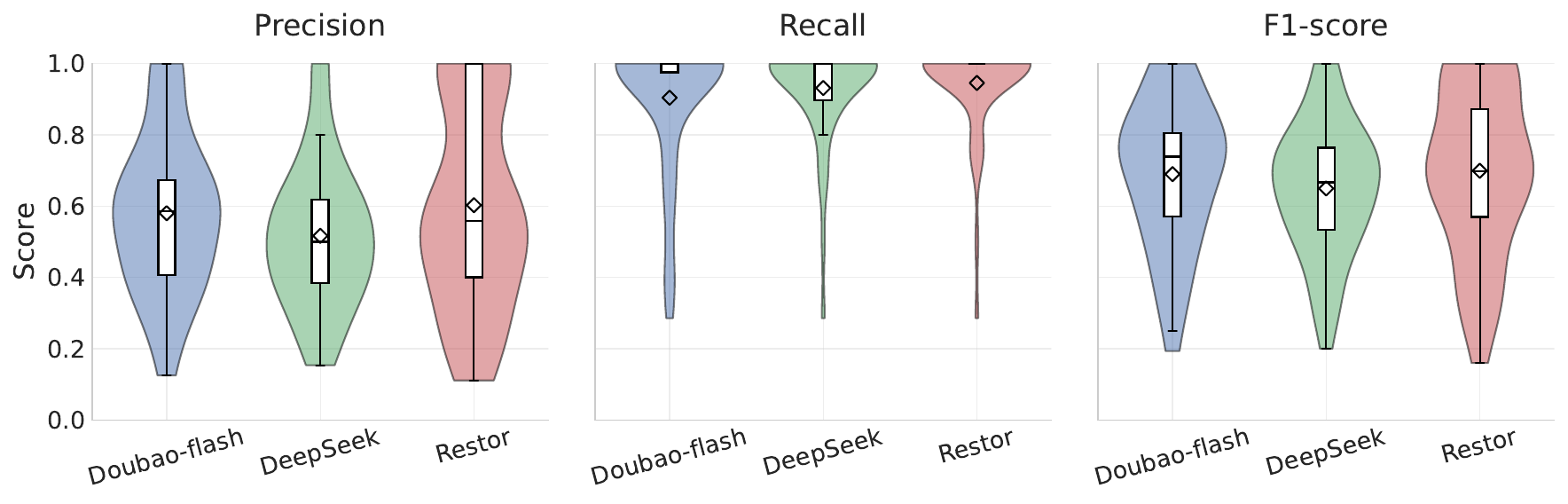}
    \Description{A composed visualization containing three subplots for precision, recall, and $F_1$~score respectively. Each subplot uses a violin plot to show the probability density of the metric across the 48 cases, overlaid with a box plot showing the interquartile range and a white diamond indicating the mean. DeepSeek shows a wide, low distribution for precision with a mean around 50\%, while \toolname shows a tighter distribution concentrated at higher values with a mean around 60\%.}
    \caption{Distribution of Key Field Identification Metrics in User Study. The figure integrates violin plots (density), box plots (quartiles), and diamond markers (mean) to visualize model performance variance on real-world data.}
    \label{fig:rq2_field_identification}
\end{figure}

\begin{figure}[h]
    \centering
    \includegraphics[width=0.9\linewidth]{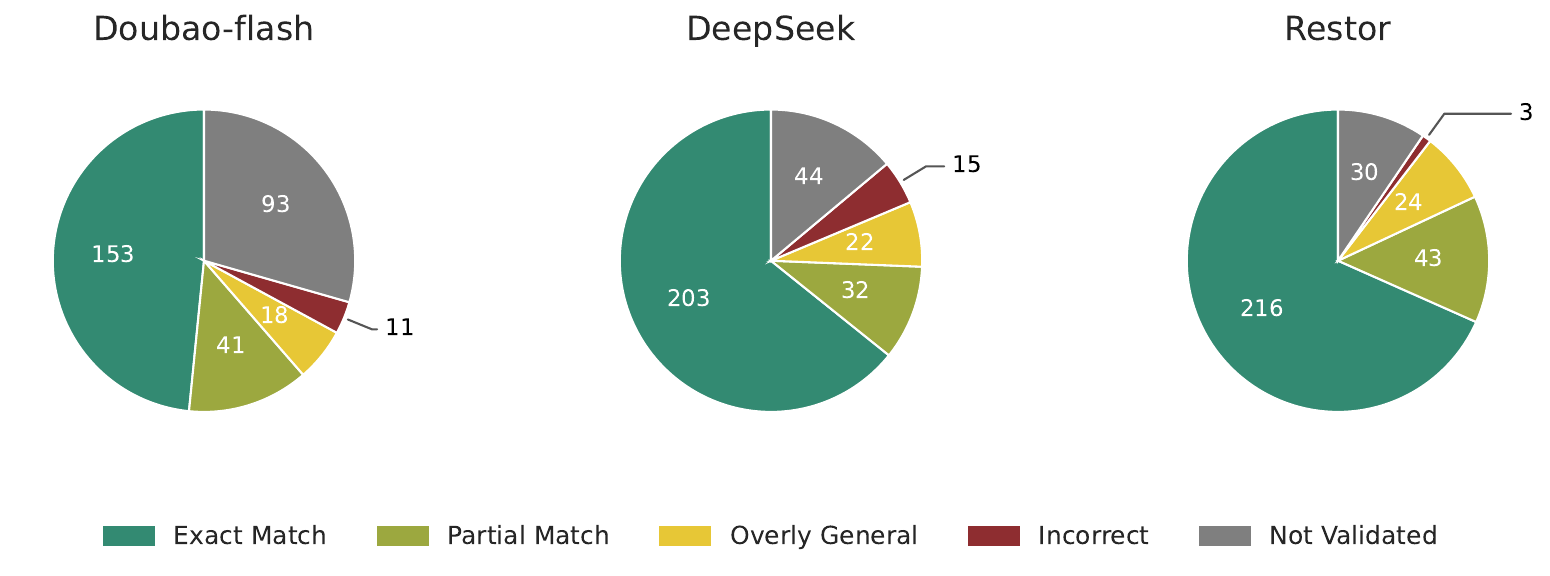}
    \Description{A visualization of assertion quality categories. It compares the total count of assertions falling into 'Exact Match', 'Partial Match', 'Overly General', 'Incorrect', and 'Not Validated' categories. \toolname has the highest number of Exact Matches (216) and the lowest number of Incorrect assertions (3). DeepSeek follows with 203 Exact Matches but higher incorrect counts. \baseModelShort lags significantly with lower matches and high missed counts.}
    \caption{Distribution of Assertion Accuracy Categories. The chart quantifies the semantic correctness of the generated logic, highlighting the volume of usable assertions (Exact Match) versus errors (Incorrect) and omissions (Not Validated).}
    \label{fig:rq2_assertion_accuracy}
\end{figure}

\subsubsection{Key Field Identification}
\label{sec:rq2_field}

Similar to \myrq{1}, we evaluate key field identification using precision, recall, and $F_1$~score against expert-reviewed key field labels.

Figure~\ref{fig:rq2_field_identification} presents the detailed distribution of these metrics. 
Specifically, \toolname achieves a precision of 60.27\%, a recall of 94.54\%, and an $F_1$~score of 69.91\%, demonstrating a concentrated distribution that suggests improved robustness on noisy industrial traffic. 
In contrast, DeepSeek yields a mean recall of 93.13\% but a lower precision of 51.67\%. 
These results indicate that while DeepSeek adopts an aggressive selection strategy that frequently includes non-deterministic fields requiring manual filtering, \toolname achieves the best overall balance between coverage and selectivity.

\subsubsection{Assertion Accuracy}
\label{sec:rq2_assert}

Beyond field selection, we evaluate whether the generated assertions match expert expectations. Figure~\ref{fig:rq2_assertion_accuracy} summarizes the semantic quality of assertions using the same five categories defined in Section~\ref{subsec:exp_setup}.

\toolname generates the largest number of \textit{Exact Match} assertions (216), exceeding DeepSeek (203) and \baseModelShort (153). Moreover, \toolname exhibits high reliability, producing only 3 \textit{Incorrect} assertions across all cases, while DeepSeek generates 15 incorrect assertions. These results indicate that \toolname yields more directly usable test logic and reduces the risk of false positives that can interrupt CI/CD pipelines.

\subsubsection{Categorized Analysis of Expert Feedback}
\label{sec:rq2_feedback}

To complement the quantitative findings, we conducted a qualitative user study involving 13 QA engineers across 11 distinct business lines. Participants were asked to evaluate the generated oracles based on four specific criteria: (1) \textit{Relevance} (whether the oracles focus on the critical fields actually monitored in production), (2) \textit{Accuracy} (the factual correctness of the assertions), (3) \textit{Experience} (satisfaction with generation speed and system latency), and (4) \textit{Open Feedback}. We categorized their insights into four recurring core themes regarding operational utility and boundary cases:

\begin{itemize}
    \item \textbf{Actionable Oracle Generation:} Experts noted that baseline models often rely on shallow structural checks, such as generic existence or type validation. In contrast, \toolname significantly reduces manual post-editing effort by producing deeper, semantically rigorous assertions.
    
    \item \textbf{High Execution Efficiency:} Most users highlighted a substantial improvement in generation speed compared to large-scale general LLMs, particularly when generating comprehensive test cases with numerous assertions. This reduction in latency enhances developer experience and facilitates seamless iterative testing within daily workflows.
    
    \item \textbf{Payload Scale Boundaries:} In a few extreme instances (3/48), the generated assertions became relatively verbose when processing complex response bodies with exceptional field cardinality ($>400$ fields), which outstrips the typical training distribution ($<100$ fields). Under such sparse edge cases, the assertion density increases, occasionally requiring minor manual refactoring to optimize long-term maintainability.
    
    \item \textbf{Context-Dependent Domain Constraints:} Certain highly specialized business rules (e.g., proprietary, complex billing calculations) could not be entirely inferred due to the deterministic black-box nature. Future work will explore incorporating lightweight domain-specific prompts during inference to address this without retraining.
\end{itemize}

\begin{tcolorbox}[colback=LightGray, title=Answer to \myrq{2}, colframe=black, left=1mm, right=1mm, top=1mm, bottom=1mm]
Expert evaluation confirms that \toolname outperforms baselines in both reliability and utility for real-world testing. Quantitatively, it achieves the highest semantic accuracy with minimal errors and a superior balance in key field identification. Qualitatively, users report significantly reduced manual editing efforts and improved generation efficiency, validating the model's effectiveness for industrial workflows despite minor limitations with extreme payload sizes.
\end{tcolorbox}

\subsection{Analysis of RQ3: Industrial Application}

To evaluate the performance of \toolname in the production environment, we monitored the system's usage on the test generation platform (introduced in Section~\ref{subsec:exp_setup}) starting from early November 2025. We analyze the \textbf{Adoption Rate} across two dimensions: temporal trends over five bi-weekly intervals and distribution across distinct business lines.

\subsubsection{Temporal Trend of Adoption Rate}
\label{sec:rq3_time}

\begin{figure}[h]
    \centering
    \includegraphics[width=0.7\linewidth]{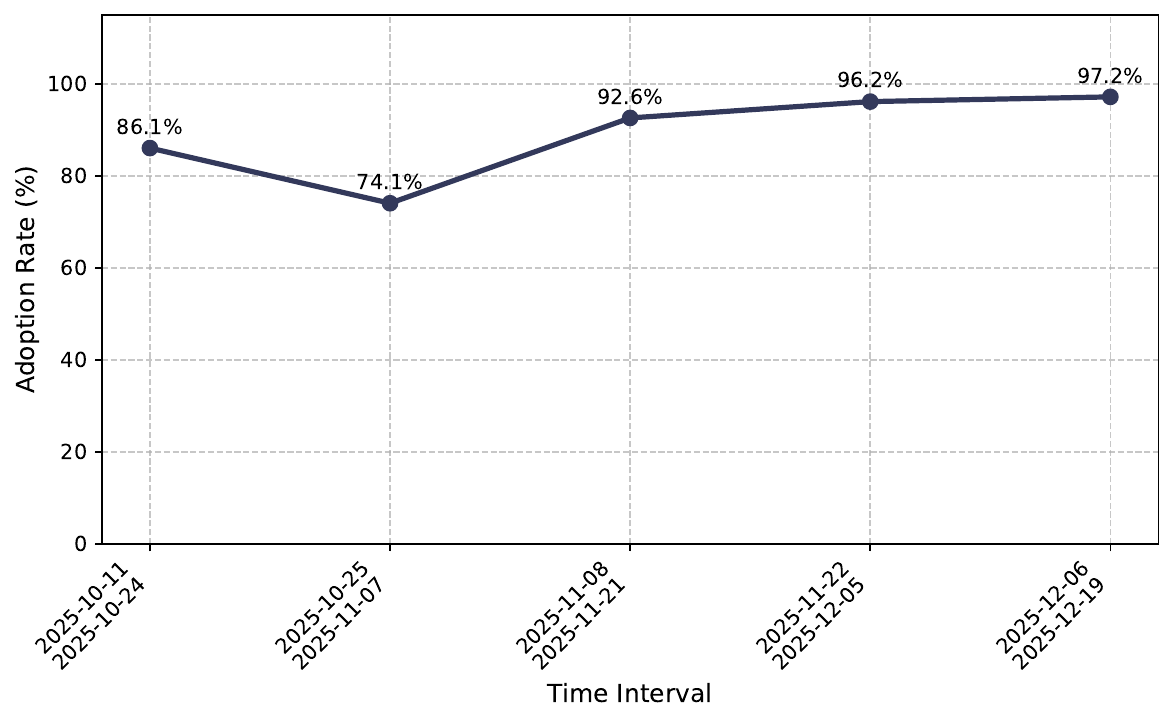}
    \Description{A line graph tracking test case adoption rates over five bi-weekly intervals. The rate starts at 86.1\%, dips to 74.1\%, and then rises steadily after early November, culminating at 97.2\% in the final observed period.}
    \caption{Temporal trends of adoption rate. The deployment of the fine-tuned model (starting early November) correlates with a sustained improvement in adoption, stabilizing above 97\% in the final interval.}
    \label{fig:rq3_by_time}
\end{figure}

To assess the impact of \toolname on testing efficiency over time, we analyzed the adoption trends across five bi-weekly intervals from mid-October 2025 to mid-December 2025. Figure~\ref{fig:rq3_by_time} illustrates the trajectory of user acceptance before and after the model deployment.

As illustrated in Figure~\ref{fig:rq3_by_time}, the adoption rate stood at 74.1\% prior to the full deployment of the fine-tuned model (before November 8). Following the rollout, the system demonstrated an immediate and substantial performance leap, with the adoption rate surging to 92.6\% in the initial post-deployment interval. This upward trajectory continued steadily, stabilizing above 96\% throughout December. This sustained high acceptance rate confirms the effectiveness of the GRPO fine-tuning, as the model consistently generates assertions that meet strict production standards, thereby minimizing the need for manual intervention.

\subsubsection{Adoption Rate Across Business Lines}
\label{sec:rq3_vc}

\begin{figure}[h]
    \centering
    \includegraphics[width=0.9\linewidth]{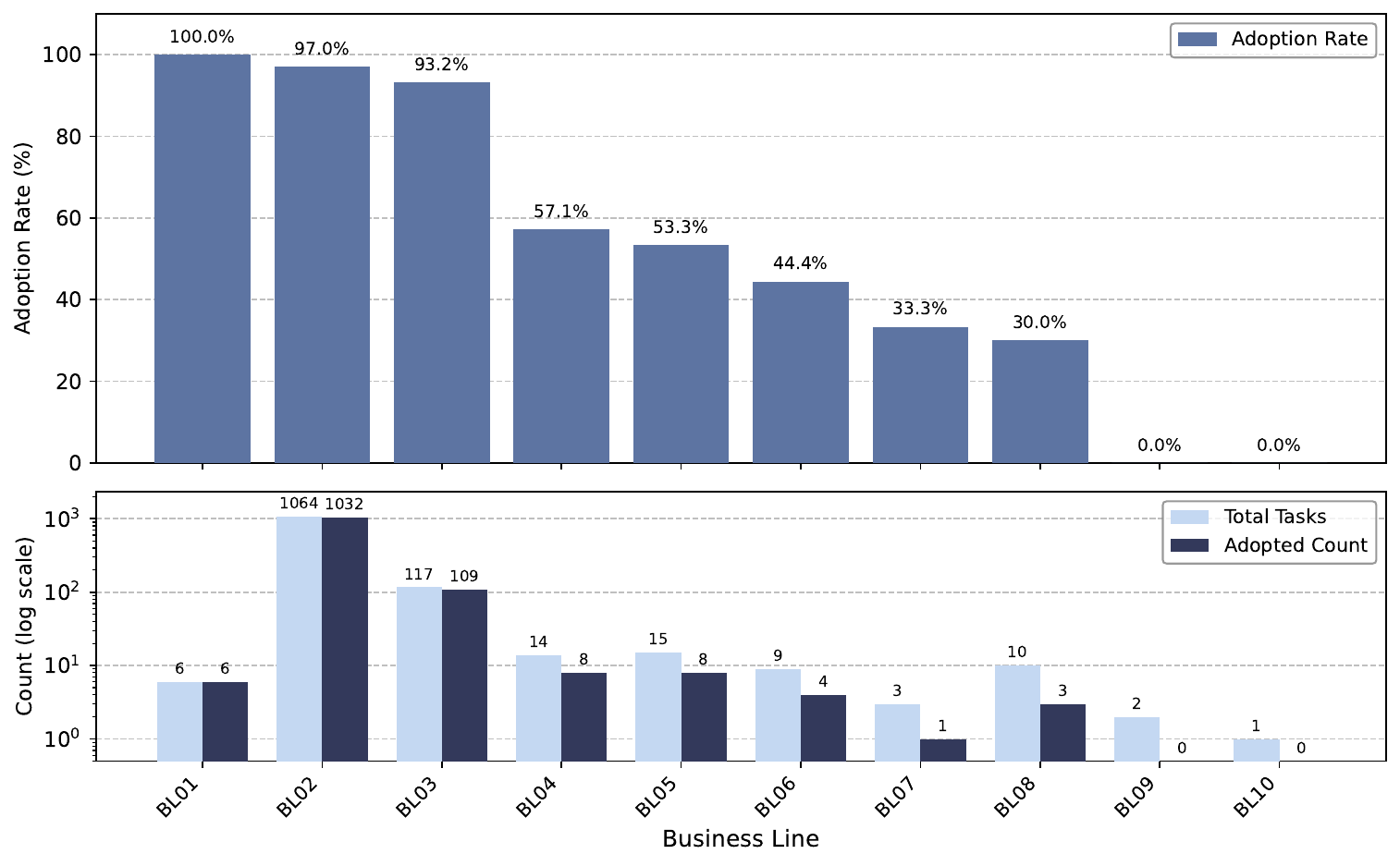} 
    \Description{A dual-axis bar chart comparing adoption rates and task volumes across ten business lines. The top chart shows adoption percentages ranging from 100\% (BL01) to 0\% (BL10). The bottom chart uses a logarithmic scale to show task counts, highlighting BL02 as the dominant contributor with 1064 tasks.}
    \caption{Adoption rates (top) and task volumes (bottom) distributed by Business Lines. Usage is concentrated in core lines (BL02, BL03) which show high adoption (>93\%).}
    \label{fig:rq3_by_vc}
\end{figure}

We further decomposed usage metrics by business line to assess domain adaptability. Figure~\ref{fig:rq3_by_vc} presents the adoption rates and generation volumes across ten business lines (BL01--BL10). A 'task' represents a single test generation request, which is initiated when a QA engineer uploads a sampled request-response traffic trace to the platform.

The results demonstrate a usage pattern concentrated in core services. BL02 and BL03, which represent the platform's primary workload, achieved adoption rates of 97.0\% (1,064 tasks) and 93.2\% (117 tasks), respectively. This confirms that \toolname successfully captures the domain logic for the system's most active users. Conversely, lines with low task volumes (BL04--BL10) exhibit higher variance. Analysis suggests this stems from two factors: specific lines with highly complex, context-heavy logic currently outside the model's scope, and new engineering teams still in the platform onboarding phase. Overall, the tool achieves high fidelity in the dominant production scenarios.

\begin{tcolorbox}[colback=LightGray, title=Answer to \myrq{3}, colframe=black, left=1mm, right=1mm, top=1mm, bottom=1mm]
    The industrial deployment confirms the practical value of \toolname. Following the model's release, the global adoption rate improved from 74.1\% to a stable level exceeding 96\%. Furthermore, in core business lines that constitute the majority of production traffic, the tool achieved an acceptance rate greater than 93\%. These results indicate that the fine-tuned model effectively delivering reliable and low-maintenance assertions for QA engineers.
\end{tcolorbox}

\section{Related Work}\label{sec:related_work}

\subsection{Automated Testing of REST APIs}
Automated testing of REST APIs primarily adopts a black-box approach, generating HTTP request sequences to maximize code coverage or uncover faults~\cite{atlidakis2019restler, kim2025llamaresttest,chen2023dynamic,zhang2024trace}. The majority of these techniques rely heavily on formal specifications, such as the OpenAPI Specification (OAS), to derive valid inputs and expected behaviors~\cite{openapi2025, martin2021restest, he2022deepstl, segura2022automated}. In the absence of specifications, tools must infer schemas from traffic, often struggling with complex dependencies.

A critical challenge in this domain is the \textit{oracle problem}. Traditional tools typically employ implicit oracles restricted to detecting crashes (e.g., 5XX status codes) or specification violations~\cite{martin2020restest, hatfield2022deriving}, failing to verify domain-specific business logic. To bridge this gap, recent works utilize either static or dynamic inference. SATORI~\cite{alonso2025satori}, a static approach, employs LLMs to deduce behavioral rules from OAS descriptions. Conversely, dynamic invariant detection methods like AGORA+~\cite{alonso2024agoraplus} mine patterns from massive execution logs. However, both paradigms have significant limitations: static methods depend on accurate, up-to-date specifications, while dynamic methods require extensive, diverse traffic to avoid overfitting. Our work targets a strictly colder start scenario—generating strict assertions from a single request-response sample without access to specifications or historical logs.

\subsection{Test Oracle Generation}
General automated oracle generation often relies on white-box or grey-box information, such as source code analysis~\cite{dinella2022toga, hossain2024togll}, method-level contracts~\cite{chen2021boosting, yu2022automated}, or rich documentation~\cite{gay2014improving}. For instance, many mature techniques operate at the method or unit level within specific programming languages (mainly Java), leveraging internal information such as dataflow, variable names, and code structure to derive assertions~\cite{chen2021boosting, dinella2022toga, hossain2024togll, watson2020learning, yu2022automated, molina2021evospex}. While effective for unit testing, these approaches are inapplicable to black-box REST API testing where only inputs and outputs are accessible.

Dynamic invariant detection remains the closest predecessor to our work for black-box systems~\cite{alonso2024agoraplus, alonso2023agora}. Yet, as noted above, its reliance on large-scale execution history makes it unsuitable for validating new features with scarce data. Recently, Large Language Models have been applied to oracle generation~\cite{he2024empirical, molinelli2025tratto}. However, most LLM-based solutions rely on prompt engineering with large models, which incurs high latency and cost, or require code context not available in black-box settings. Ours is distinct in leveraging Reinforcement Learning to fine-tune a lightweight model. This enables the agent to internalize testing semantics and generate executable assertions efficiently, bypassing the need for huge contexts or expensive general-purpose models.

\subsection{Reinforcement Learning in Test Generation}
Reinforcement Learning (RL) has proven effective in software testing by training agents to maximize objectives like code coverage or crash diversity. RL-based test generation has been successfully applied across diverse domains, including the validation of complex fundamental systems (e.g., compiler testing~\cite{chen2023compiler}, cyber-physical systems~\cite{zhang2021figcps}), domain-specific applications such as autonomous driving and robotics~\cite{lu2022learning, humeniuk2024reinforcement, doreste2024adversarial,chen2023enhancing,wu2024synthesizing}, and general-purpose softwares like Android
applications~\cite{pan2020reinforcement,cai2024reproducing,guo2022detecting,sun2021understanding,dong2020time} and CI systems~\cite{nouwou2023comparison}. In these contexts, RL agents are typically trained to generate effective test inputs or environment configurations that maximize a calculated reward, such as crash probability~\cite{lu2022learning} and coverage~\cite{humeniuk2024reinforcement}. The central objective in these works remains optimizing the sequence or structure of the inputs fed to the system under test to achieve higher fault detection.

% While the application of RL to test input generation is mature, its use in solving the test oracle problem remains highly limited. For instance, the recent AutoRestTest framework applies Multi-Agent Reinforcement Learning (MARL) along with LLMs to the REST API domain~\cite{kim2024multi}. Crucially, however, the goal of AutoRestTest is dependency-aware test sequence generation to improve coverage, reinforcing the existing focus on input-side optimization. To the best of our knowledge, our work represents the first application of Reinforcement Learning to the specific challenge of automated test oracle generation for black-box REST APIs. We leverage RL to fine-tune a pre-trained LLM, enabling it to learn a robust policy for transforming a single, raw request-response pair into high-quality, executable oracle code, thereby tackling the oracle problem under the severe constraints of unavailable specifications and single-instance traffic data.

Critically, existing RL-based testing works focus almost exclusively on \textit{input generation}. The application of RL to the \textit{oracle problem}, i.e., verifying the correctness of the output, remains unexplored. To the best of our knowledge, our work represents the first application of RL specifically for generating test oracles for REST APIs in a black-box context. Instead of optimizing input fuzzing, we train the model to distinguish strict logic boundaries, transforming raw responses into high-quality, executable verification code.

\section{Threats to Validity}
\label{sec:threats}

We define potential threats to the validity of our study and discuss the mitigation strategies employed to address them.

\subsection{Internal Validity}

\paragraph{\textbf{Human Annotation Bias}}
The reliance on human judgment for test oracle validation (\myrq{1} and \myrq{2}) introduces risks of subjectivity and fatigue. To mitigate this, we employed three senior QA engineers from \companyname. We implemented a strictly blinded consensus mechanism, where ground truth labels were established only upon majority agreement. This approach minimizes individual bias and ensures the evaluation aligns with objective industrial standards.

\paragraph{\textbf{Data Leakage}}
To ensure the model learns generalized logic rather than memorizing training data, we adopted a two-tiered strategy. First, for \myrq{1}, we enforced strict contamination control by physically excluding test set API endpoints from the training corpus. Second, the real-world evaluation in \myrq{2} utilizes traffic from \textit{newly developed} features not yet integrated into the traffic recording platform. Consequently, these samples are inherently disjoint from the training distribution, effectively precluding data leakage.

\paragraph{\textbf{Validity of Adoption Rate Metric}}
In \myrq{3}, we use adoption rate as a proxy for utility. A potential threat is that users might commit generated code without sufficient scrutiny. However, the deployment workflow at \companyname mandates peer code review. Therefore, committed assertions represent code that has satisfied not only the author but also independent human reviewers, validating the practical quality of the generation.

\subsection{External Validity}

\paragraph{\textbf{Generalizability of Source Data}}
Our dataset originates entirely from \companyname's production environment, posing a risk of overfitting to specific corporate conventions (e.g., specific envelope structures or naming schemes) that may not apply to other organizations. To proactively mitigate this threat and ensure broader generalizability, we intentionally sampled data across 15 distinct business lines and 246 diverse services. This extensive variety encompasses a wide spectrum of RESTful design patterns, payload complexities, and data logic prevalent in the broader software industry. While the deployed model may inevitably learn certain internal formatting conventions, the underlying reinforcement learning methodology for internalizing oracle generation logic is highly generalizable. We assert that this framework can be readily transferred to other organizational contexts and API paradigms (such as RPC) with appropriate data collection and constraint definitions.

\paragraph{\textbf{Language Specificity}}
The current implementation of \toolname generates test assertions exclusively in Python. This limits the direct application of our fine-tuned model to environments using other technology stacks (e.g., Java/JUnit or JavaScript/Jest). However, the core contribution of this work lies in the reinforcement learning methodology that aligns LLMs with testing logic ("common sense"), rather than language syntax. We believe the proposed approach can be readily adapted to other programming languages by substituting the ground truth samples and execution feedback mechanisms in future work.

\section{Conclusion}\label{sec:conclusion}

In this paper, we presented \toolname, a reinforcement learning-driven framework designed to automate the generation of precise semantic test assertions for REST APIs. By adapting a lightweight Large Language Model via GRPO, our approach effectively internalizes domain-specific testing logic, enabling the system to identify business-critical fields and infer strict logical constraints from single traffic samples without relying on formal specifications. Extensive experiments demonstrate that \toolname significantly outperforms both prompt-engineered baselines and large-scale general models (e.g., DeepSeek\cite{deepseek2025terminus}\cite{deepseekai2024deepseekv3technicalreport}) in terms of key field identification precision and assertion safety. Furthermore, the successful deployment of \toolname within \companyname's production testing platform—achieving a sustained adoption rate exceeding 90\% across core business lines—validates its practical utility and robustness in accelerating industrial testing workflows.

%% Appendix and other contents

\begin{acks}
{\sloppy This work was supported by the National Key R\&D Program of China under Grant No. 2024YFB4505902.\par}
\end{acks}

\appendix

\section{Prompt Templates}
\label{appendix:prompt}

\begin{promptbox}{System Prompt Template for Test Oracle Generation}

\hlyellow{Role} \\
You are a professional API testing expert. You are skilled at generating high-quality assertions based on limited API information with your professional knowledge.

\hlyellow{Task} \\
Your task is to analyze the relevant information of an API, combine it with some historical traffic, and generate assertions for this API.

\hlyellow{Instructions} \\
Please generate assertions implemented in Python. The overall requirements are as follows: \\
\textbullet~ Based on the API information and your understanding, select important fields from the response body to assert. \\
\textbullet~ Assertion types include value equality, value range checks, object field existence checks, etc. \\
\textbullet~ The assertions you generate should be applicable to all scenarios of this API, not limited to the specific response body provided in the example traffic. \\
\textbullet~ Generate Python assertion statements only.

\hlyellow{Notes} \\
\textbf{For selecting assertion fields, it is recommended to:} \\
\textbullet~ Focus on the core resources expected to be returned by the API, for example, \texttt{users} in the response body of \texttt{getUsers}. \\
\textbullet~ Choose fields with clear semantic constraints, such as URIs (format validation), counters (non-negative numbers), etc. \\
\textbullet~ DO NOT consider: timestamps or other random fields, meaningless fields.

\textbf{For assertions on different types of fields, after understanding the field semantics and inferring their constraints:} \\
\textbullet~ Perform format validation for strings, range validation for numbers. \\
\textbullet~ For enum-like fields, you may further perform set membership validation. \\
\textbullet~ If it is truly difficult to determine constraints, only provide type validation.

\hlpink{Input and Output Format} \\
...

\end{promptbox}

% See https://conf.researchr.org/track/issta-2026/issta-2026-research-papers#open-science-policy-and-data-availability-section
\section*{Data Availability}

The research presented in this paper was conducted within \companyname, utilizing proprietary industrial datasets and internal closed-source models. Due to strict corporate non-disclosure agreements and information security policies regarding user privacy and intellectual property, the source code, raw data, and trained models cannot be made publicly available.

%%
%% The acknowledgments section is defined using the "acks" environment
%% (and NOT an unnumbered section). This ensures the proper
%% identification of the section in the article metadata, and the
%% consistent spelling of the heading.
% \begin{acks}
% To Robert, for the bagels and explaining CMYK and color spaces.
% \end{acks}

%%
%% The next two lines define the bibliography style to be used, and
%% the bibliography file.
\bibliographystyle{ACM-Reference-Format}
\bibliography{citations}

\end{document}